\documentclass[%
 aip,
 amsmath,amssymb,
 reprint,%
]{revtex4-1}

\usepackage{graphicx}
\usepackage{dcolumn}
\usepackage{bm}

\usepackage[utf8]{inputenc}
\usepackage[T1]{fontenc}
\usepackage{mathptmx}
\usepackage{etoolbox}

\makeatletter
\def\@email#1#2{%
 \endgroup
 \patchcmd{\titleblock@produce}
  {\frontmatter@RRAPformat}
  {\frontmatter@RRAPformat{\produce@RRAP{*#1\href{mailto:#2}{#2}}}\frontmatter@RRAPformat}
  {}{}
}%
\makeatother
\begin{document}

\preprint{AIP/123-QED}

\title[A synergistic approach to optical modeling of PCSELs]{A synergistic approach to optical modeling of PCSELs   \\ through rigorous methods and the coupled-wave theory }
\author{A. Bisianov}
\author{A. Waag}%
 \email{a.bisianov@tu-braunschweig.de}
\affiliation{Institute of Semiconductor Technology, Technische Universität Braunschweig, Braunschweig 38106, Germany
}%
\affiliation{Nitride Technology Center (NTC), Technische Universität Braunschweig, Braunschweig 38106, Germany}%


\date{\today}

\begin{abstract}
A wide range of numerical and semi-analytical approaches has been developed for optical modeling of photonic-crystal surface-emitting lasers (PCSELs). However, a systematic framework for comparing their predictive capabilities and identifying their respective validity limits remains largely unexplored. In this work, we introduce a comparative methodology in which four representative methods—including rigorous numerical and effective-index-based approaches—are analyzed and partially hybridized within a coupled-wave-theory framework. Using single- and double-lattice PCSELs as representative models, we demonstrate that this approach not only reveals fundamental differences between predictions of rigorous methods and the coupled-wave-theory framework, but also captures leaky and symmetry-broken phases of bound states in the continuum (BICs) relevant for laser operation.
\end{abstract}

\maketitle

\section*{\label{sec:level1}Introduction}

First proposed in 1999 \cite{imada1999coherent,meier1999laser}, photonic-crystal surface-emitting lasers (PCSELs) have evolved into a rapidly advancing chip-scale technology\cite{noda2023high,ishizaki2019progress,pan2024recent,noda2024photonic,sakata2023photonic,katsuno2024design,ishimura2023proposal,huang2025unveiling}. They continue to push the performance limits of conventional semiconductor lasers, particularly in terms of brightness, beam quality, and emission directionality. 

Alongside these technological advances and the growing diversity of PCSEL platforms, a wide range of electro-optical design approaches and numerical tools has been developed. Among the most established semi-analytical methods is the three-dimensional coupled-wave theory (3D-CWT), which enables the analysis of both infinitely periodic and finite PCSEL structures and was originally formulated in Refs.\cite{liang2011three,liang2012three,liang2013three,peng2012three}. Subsequently, this framework was extended to more complex and optically powerful photonic-crystal geometries, most notably the double-lattice photonic crystal (PhC)\cite{yoshida2019double,yoshida2023high,emoto2022wide,inoue2022general,de2020thermal}. Time-dependent version of 3D-CWT, incorporating carrier–photon interactions and nonuniform temperature profiles, has also been developed and applied\cite{inoue2019comprehensive,katsuno2021self,morita2024demonstration,morita2024high}. Beyond PCSEL designs based on highly symmetric nanostructures, such as cylindrical air holes, 3D-CWT has also been employed to investigate PCSELs with more asymmetric geometries, including triangular holes\cite{hirose2014watt,king2024design,radziunas2024optical}, rotatable double-ellipse patterns\cite{ke5400511design}, and structures with tilted sidewalls \cite{peng2011coupled} or spatially varying hole depths\cite{itoh2024high,lang2025mode, yoshida2019double}.

Despite its widespread adoption and versatility, the semi-analytical 3D-CWT is not the only numerical approach developed in recent years for PCSEL design. For instance, several studies have employed rigorous coupled-wave analysis (RCWA)\cite{xu2024modeling,song2018first,liu2023design}, which enables accurate calculations of resonant frequencies, quality factors, and optical field distributions, as well as related laser performance metrics. 
More recently, Persson et al. introduced the k-space weighted loss estimation (kSWLE) approach\cite{persson2025finite}, which allows for accurate evaluation of both vertical and lateral optical losses of lasing modes and thus provides an alternative to finite-size implementations of 3D-CWT. Loss estimation in finite PCSELs has also been addressed using a probabilistic Markov chain model, as proposed in Ref.~\cite{liu2023probabilistic,liu2025interdependence}. Another study by Lang et al.\cite{lang2025assessment}, which is particularly relevant to the present work, compares several effective-index-based approaches, including the iterated effective-index method (I-EIM), with the semi-analytical guided-mode expansion (GME) method\cite{zanotti2024legume} and rigorous methods such as RCWA and the transmission-line method (TLM). 

In addition, fully three-dimensional simulations based on the finite-element method (FEM)—as implemented in COMSOL Multiphysics—have also been applied to PCSELs, for example, in Ref.~\cite{hsieh2025impact}.

Although both effective-index-based and fully three-dimensional methods have been extensively developed and applied to PCSELs, a systematic comparative analysis that identifies qualitative differences in their predictions and explores their potential for predictive synergy remains lacking. In this article, we introduce such a comparative scheme, wherein rigorous numerical methods are employed as "phenomenological" inputs within a generalized coupled-wave theory framework. Since this framework incorporates explicit input from a chosen rigorous method, it can also be regarded as a hybridized CWT framework. While the proposed technique is, in principle, applicable to a wide variety of PCSEL types, it is demonstrated here using two otherwise well-studied configurations. 

Specifically, this article compares two rigorous three-dimensional methods—namely the finite-element method (FEM) and rigorous coupled-wave analysis (RCWA), also known as the Fourier modal method (FMM)—with two effective-index-based approaches, including the widely used 3D-CWT and the recently proposed iterated effective-index method (I-EIM). These methods are applied to blue gallium nitride (GaN)-based PCSELs, closely resembling watt-class devices reported in Ref.\cite{emoto2022wide}. Two PCSEL variants operating at $\lambda\approx 430$ nm are considered: structures incorporating subwavelength-thin photonic-crystal (PhC) layers with either single-lattice or double-lattice patterns of circular air holes, embedded in bulk GaN. In both cases, the PhC patterns are arranged on a periodic square lattice with a period of $a=174$ nm, as illustrated in Figs.~\ref{fig:lattices}(a,b). Despite our focus on these specific square-lattice types, the comparative framework introduced here can be readily extended to a broader class of lattice types and PhC patterns. 

The remainder of this article is organized as follows. In the first section, we introduce the two PCSEL configurations in detail. The subsequent section is devoted to the methodological aspects of the numerical approaches considered. As will become evident, all four methods provide full or partial access to key optical characteristics, including resonant frequencies and quality factors of transverse-electric (TE) modes relevant for lasing operation, as well as the associated photonic-crystal (PhC) confinement factors. In addition, microscopic parameters governing in-plane and out-of-plane optical feedback are extracted from FEM, FMM, and, to a limited extent, I-EIM by invoking a generalized coupled-wave theory (CWT)\cite{inoue2022general,liang2011three,liang2012three}. In contrast, within 3D-CWT, these quantities are derived directly from first principles. In this way, we enable a comparison of rigorous and effective-index-based methods on a common basis within the generalized CWT framework. In the following section, we present numerical results for PCSEL platforms featuring two PhC pattern variations and discuss the relative accuracy of each method, along with the validity limits of the generalized CWT framework. The final section summarizes the main findings.

 \section{\label{sec:level1}Two PCSEL models and BIC states}
\begin{figure}
\includegraphics[width=0.24\textwidth]{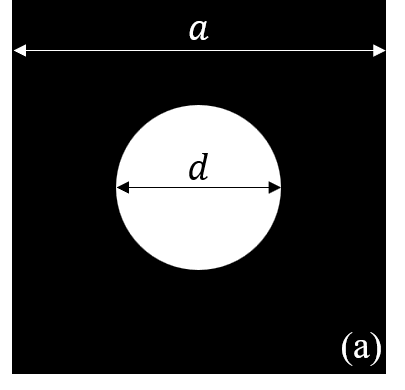}
\includegraphics[width=0.24\textwidth]{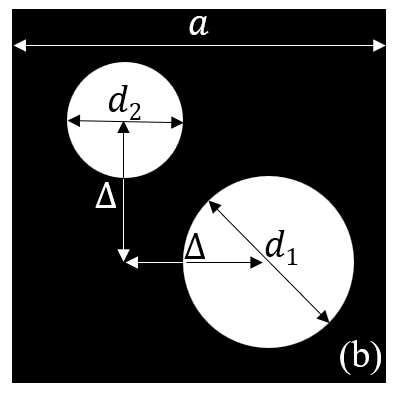}
\caption{\label{fig:lattices}Refractive-index profile of a unit cell of the single-lattice (left) and double-lattice (right) photonic crystals. White regions correspond to air holes, while black regions denote bulk GaN.}
\end{figure}

In this section, we introduce the two aforementioned types of square lattice as illustrated in Fig.~\ref{fig:lattices}(a,b). The simplest single-lattice, which features one cylindrical air hole of diameter $d$ per lattice unit cell, will be further referred to as Type I. The PCSEL as a whole and the PhC layer, incorporating such a pattern, will be equally called Type I later on. By definition, air hole filling factor $ff$  is defined via the hole diameter as $\pi d^2/(4a^2)$, where $a$ is the lattice period of the square unit-cell $a\times a$. We fix $a$  to 174 nm throughout this study, thus tuning the second-order Bragg resonance wavelength $\lambda_{\text{Bragg}}\approx a*n_{\text{eff}}$ to around 430 nm. An optimal PhC thickness $t_{\text{PhC}}$ lies in the order of half-wavelength in the material, which is $\lambda_{\text{Bragg}}/(2n_{\text{eff}}) \sim100$ nm. Therefore, the thickness is fixed to 100 nm in both PCSEL types. Additionally, we probe a variety of PhC thicknesses in the Type I case.

Next, a basic double-lattice PhC features two cylindrical holes of diameters $d_1$ and $d_2$, respectively. For convenience, we refer to this pattern as Type II. The corresponding hole filling factors are $ff_1$ and $ff_2$. The lateral distance between the holes along the $x$ and $y$ axes is defined as $\Delta=\Delta_x=\Delta_y$ as shown in Fig.~\ref{fig:lattices}(b). In our parametric study, we ensure that $\Delta>(d_1+d_2)/(2\sqrt{2}) $, thus preventing the two holes from merging.

To make the two PCSEL types comparable, both Type I and Type II PhC layers are embedded in nearly identical GaN-based epitaxial multilayer stack, as summarized in Table~\ref{tab:table1}.  The only distinction between the two designs is the refractive index of the n-type optical cladding, which is intentionally set higher in the Type I structure to promote observable mode leakage through that cladding already at moderate hole diameters, as demonstrated below. 

As already mentioned, the multilayer stack under study closely resembles blue PCSELs realized in \cite{emoto2022wide}. The presented PCSEL platform has not been intentionally optimized as an ultimate design proposal and serves here as merely a realistic study model. Neither the optical losses associated with material absorption nor the optical gain of the active region are included in our model. Additionally, the top p-type electrode pad has been omitted since the ITO layer of 200 nm proved to be sufficiently thick to optically isolate fundamental guided modes of interest. Having said that, it is important to keep in mind that a p-type electrode can act as an optical reflector, which redirects and redistributes the optical field, thus leading to a significant modification of upward and downward radiation rates $\alpha_{\text{up,down}}$ of a lasing device (on that matter see e.g. \cite{emoto2022wide,ishizaki2023enhancement,lang2025mode}).  Here, we do not take into account or make an accent on such effects, since power maximization and laser performance optimization are beyond the scope of this study. Moreover, we do not take into account any finite-size effects, such as, e.g., in-plane optical losses, assuming a  periodically infinite PCSEL in the lateral $xy$-plane.

Throughout this study, we focus on four fundamental quasi-TE guided resonance modes common to both Type I and Type II PhC patterns and well established in the literature\cite{inoue2022general,liang2011three,liang2011three,noda2023high}, as the most relevant modes in standard PCSELs with TE-polarized optical gain. These modes and their associated photonic bands are conventionally denoted as A, B, C, and D, with their fourfold multiplicity originating from the $C_4$ rotational symmetry of the underlying square lattice. In a broader context of photonic-crystal theory, band-edge eigenstates A and B, located at the second-order $\Gamma$ point of the photonic band structure, are classified as at-$\Gamma$ symmetry-protected bound states in the continuum (BIC)\cite{hsu2016bound,wang2024optical,zhao2023large,li2025observation,yang2014analytical}.  These states exhibit infinite quality factors, as they do not couple to free-space radiation owing to their symmetry mismatch with existing radiative channels. In theory, symmetry-protected BICs are immune to perturbations preserving the underlying symmetry of the lattice, which is the case for the Type I lattice with one circularly symmetric hole. In contrast, the two holes of the Type II lattice break the $C_4$ rotational symmetry of the square lattice, thus lifting the symmetry protection and forcing the band-edge BIC states A and B to radiate at the second-order $\Gamma$ point. In real devices, fabrication-induced distortions of the PCSEL stack geometry, particularly of the PhC hole pattern, render the Q factors of modes A and B finite, even in the Type I lattice. In this case, the more precise term “quasi-BIC state” is used to account for a high yet finite Q factor of the mode. As will be demonstrated in the Type I PCSEL study below, rigorous numerical methods likewise yield finite Q factors for at-$\Gamma$ BIC states for two independent reasons: (i) finite mesh discretization of the ideal circular hole shape, which inevitably breaks the symmetry protection (but can be mitigated by further mesh refinement), and (ii) the proximity of the BIC effective index to that of the optical cladding, which results in weak mode confinement and leakage.

\begin{table}
\caption{\label{tab:table1}Layer properties of the modeled PCSEL stack, featuring either single-lattice (Type I) or double-lattice (Type II) PhC.}
\begin{ruledtabular}
\begin{tabular}{lcr}
Layer material &Thickness [nm] & Refractive index\\
\hline
air & —  & 1\\
ITO & 200 & 2.02\\
p-GaN & 25 & 2.5\\
p-AlGaN& 400 & 2.47\\
p-AlGaN & 15 & 2.42\\
ud-GaN & 130 & 2.5\\
ud-InGaN & 3 & 2.73\\
ud-GaN & 6 & 2.5\\
ud-InGaN & 3 & 2.73\\
ud-GaN & 6 & 2.5\\
ud-InGaN & 135& 2.51\\
PhC & 100& 2.5 (bulk), 1 (hole)\\
n-GaN & 80 & 2.5\\
n-AlGaN& 2000 & 2.48 (Type I), 2.47 (Type II)\\
n-GaN substrate & — & 2.5

\end{tabular}
\end{ruledtabular}
\end{table}
\section{\label{sec:level1}Methodology}


In this section, we describe four methods applied to the earlier introduced PCSEL configurations. The first pair of methods is the aforementioned finite element method (FEM) and the rigorous coupled-wave analysis (RCWA), which is also called the Fourier modal method (FMM). They are implemented via commercially-available COMSOL Multiphysics and Ansys Lumerical software, respectively. The second pair of methods is the semi-analytical effective-index-based 3D-CWT\cite{liang2011three,liang2012three,inoue2022general}, and I-EIM\cite{lang2025assessment} approaches, introduced earlier in the text. In contrast to rigorous methods, which evaluate electromagnetic fields as fully three-dimensional entities, the semi-analytical ones effectively homogenize the refractive index profile of the PhC layer, thereby enabling a separable representation of the electric field of the form $E_{\text{1D}}(z)E_{\text{2D}}(x,y)$.  One of the main drawbacks of the 3D-CWT method is the assumption that the four fundamental quasi-TE modes share an identical vertical field profile $E_{\text{1D}}(z)$. As will be shown below, the I-EIM can, to some extent, mitigate this limitation. By contrast, FEM and FMM do not rely on this assumption and instead fully resolve the three-dimensional geometry of the PCSEL structure.

Despite these simplifying assumptions, 3D-CWT remains a valuable and instructive approach, as it enables an analytical derivation of so-called Hermitian $\hat{C}_{\text{1D}}, \hat{C}_{\text{2D}}$ and non-Hermitian $\hat{C}_{\text{rad}}$ field coupling matrices, governed by the associated coupling coefficients.  In accordance with Ref.~\cite{inoue2022general}, we utilize the following CWT-based system of nonlinear equations, which we hereafter refer to as the CWT framework:
\[
\delta_{\text{A,C}} + i \alpha_{\text{A,C}}/2
= \kappa_{11} + \kappa_{2\mathrm{D}+} + i\mu\tag{1}\label{eq1}\]
\[\mp\sqrt{
\left[(\kappa_{1\mathrm{D}} + \kappa_{2\mathrm{D-}}) + i\mu e^{i\theta_{\text{pc}}}\right]
\left[(\kappa_{1\mathrm{D}} + \kappa_{2\mathrm{D-}})^{*} + i\mu e^{-i\theta_{\text{pc}}}\right]
}, 
\]
\[
\delta_{\text{B,D}} + i \alpha_{\text{B,D}}/2
= \kappa_{11}- \kappa_{2\mathrm{D+}} + i\mu\tag{2}\label{eq2}\]\[\mp \sqrt{
\left[(\kappa_{1\mathrm{D}} -\kappa_{2\mathrm{D-}}) + i\mu e^{i\theta_{\text{pc}}}\right]
\left[(\kappa_{1\mathrm{D}} - \kappa_{2\mathrm{D-}})^{*} + i\mu e^{-i\theta_{\text{pc}}}\right]
}.
\]
These equations explicitly relate the complex eigenvalues of the fundamental modes to the coupling coefficients that govern internal field dynamics and wave interactions. Here, $\delta_{i}$ denote deviations of the corresponding spatial eigenfrequencies from the Bragg wavenumber $\beta_0=2\pi/a$, which is conveniently expressed in $\text{cm}^{-1}$. The quantities $\alpha_{i}=\beta_0/Q_{i}$ represent the radiation constants associated with the respective modal quality factors. The in-plane distributed feedback of a PCSEL is governed by the coupling coefficients $\kappa_{\text{1D}}$ and $\kappa_{\text{2D}\pm}$, which play a central role in determining lasing performance and optical losses. In this study, we therefore focus primarily on these parameters. Specifically, we consider their absolute values, as these directly influence key laser performance metrics such as in-plane losses and far-field beam shape, whereas the phases of the generally complex coupling coefficients are typically less transparent and relevant for performance optimization.

Within the FEM and FMM methods, the spatial eigenfrequencies $\overline{\nu}_{i}=\beta_0+\delta_{i}$ and the associated radiation constants $\alpha_{i}$, are estimated as described in the corresponding subsections below. By contrast, our I-EIM formulation does not provide access to radiation constants for reasons discussed in the methodology section; consequently, only real-valued eigenfrequencies are obtained using this method. 

For Type II PCSEL, only the real-valued coupling coefficient $\kappa_{\text{2D+}}$ can be estimated based on I-EIM through:
\[ \kappa_{\text{2D+}}=(\overline{\nu}_{\text{A}}-\overline{\nu}_{\text{B}}+\overline{\nu}_{C}-\overline{\nu}_{D})/4\tag{3}\label{eq3}\]In case of Type I PhC, 3D-CWT yields $\theta_{\text{pc}}=\pi$ and $\alpha_{\text{A,B}}=0$, which reduces Eqs.~(\ref{eq1},\ref{eq2}) to 
\[
\delta_{\text{A,B}} 
= \kappa_{11}\pm2\kappa_{2\mathrm{D}} +\kappa_{\text{1D}}, \alpha_{\text{A,B}}=0,\tag{4}\label{eq4}\]
\[
\delta_{\text{C,D}} 
= \kappa_{11}- \kappa_{\text{1D}},  \alpha_{\text{C,D}}=4i\mu \tag{5}\label{eq5}\]
and renders all coupling coefficients real-valued: 
\[\kappa_{\text{1D}}=(\overline{\nu}_{\text{A}}+\overline{\nu}_{\text{B}}-\overline{\nu}_{C} -\overline{\nu}_{D})/4,\tag{6}\label{eq6}\]
\[\kappa_{\text{2D+}}=\kappa_{\text{2D-}}=(\overline{\nu}_{\text{A}}-\overline{\nu}_{\text{B}})/4. \tag{7}\label{eq7}\]

Therefore, besides 3D-CWT, I-EIM can also provide values for all these coefficients based on the associated real-valued eigenfrequencies. It should be emphasized that, according to the rigorous FEM and FMM, the condition $\alpha_{\text{A,B}}=0$ must not be imposed a priori, as it generally breaks down at larger hole diameters, as will be demonstrated below. Accordingly, for these methods, we employ the full form of the CWT framework given by Eqs.~(\ref{eq1},\ref{eq2}).

Next, within the CWT treatment\cite{inoue2022general} of the Type II PhC, we define $\theta_{\text{pc}}=\pi+2 \text{arg}(\xi_{1,0})$, where $\xi_{j,k}$ denote Fourier harmonics of the dielectric constant profile $\varepsilon_{\text{PhC}}(x,y)$ of the PhC. As a consequence, the system of Eqs.~(\ref{eq1},\ref{eq2}) becomes overdetermined, comprising eight real-valued equations (for the real and imaginary parts of the eigenfrequencies) but only seven unknowns. 
To resolve this issue, we fix the real-valued eigenfrequencies together with the radiation constants  $\alpha_{\text{C,D}}$, while allowing the CWT framework to fit the remaining pair $\alpha_{\text{A,B}}$. This choice is not arbitrary, as the description of the BIC modes constitutes the key issue distinguishing predictions of rigorous methods from those of the CWT framework, as discussed in the following.  At a heuristic level, this choice is further motivated by the observation that the 3D-CWT-based values of $\alpha_{\text{C,D}}$ generally exhibit better agreement with their counterparts based on FEM and FMM than the corresponding values of $\alpha_{\text{A,B}}$. 

Finally, it is worth emphasizing that FEM- and FMM-based eigenfrequencies serve as a "phenomenological" input for the generalized CWT framework (Eqs.~(\ref{eq1},\ref{eq2})), while 3D-CWT, being inherently constructed within the CWT framework, derives all coupling coefficients from first principles. The following subsections describe some specific aspects of each applied method. 
 \subsection{\label{sec:level2}Three-dimensional coupled-wave theory}
The core algorithm of 3D-CWT is implemented as code programmed in Wolfram Mathematica Software. The vertical electric field profile $E_{\text{1D}}(z)$ of the factorized TE-polarized electric field ${ \mathbf{E}= \mathbf{E}_\text{2D}(x,y)E_{\text{1D}}(z)}$ is obtained via the Finite Difference Eigenmode (FDE) solver provided by Ansys Lumerical Software. Note that $\mathbf{E}_\text{2D}(x,y)$ is an in-plane vector field, corresponding to the TE polarization. The highest-order Fourier component of the PhC refractive-index profile is set to 11, resulting in a total of 22×22 superimposed Bloch waves in reciprocal space, plus the zero-order radiative waves, which act as vertically radiative channels for the guided resonant modes. This number of harmonics ensures a decent numerical precision combined with an acceptable computational time, which typically does not exceed 30 min per run. Within the 3D-CWT method, radiative and high-order waves have been derived by applying the generalized or simplified Green functions described in \cite{peng2011coupled,liang2011three,liang2012three}. The use of simplified Green functions not only significantly reduces the computational cost but also avoids the emergence of nonphysical field divergences at relatively large PhC thicknesses, as will become evident for the Type I PCSEL. However, at least for a PhC thickness of 100 nm, the generalized Green functions do not exhibit a significant divergence; consequently, the generalized ansatz is employed within 3D-CWT for this thickness in both Type I and Type II PCSELs.

\subsection{\label{sec:level2}Iterated effective-index method}
The second effective-index-based method is the I-EIM, introduced for PCSELs in Ref.~\cite{lang2025assessment}. The approach starts from the three-dimensional frequency-domain wave equation:
\[
\nabla \times \nabla \times \mathbf{E}(\mathbf{r}) = \left(\frac{\omega}{c}\right)^2 \epsilon(\mathbf{r}) \mathbf{E}(\mathbf{r}),\tag{8}\label{eq.11}\]
where $\mathbf{r}$=($x,y,z$), $c$ is the speed of light in vacuum, $\epsilon(\mathbf{r})$ is the spatially varying dielectric constant of the PCSEL structure, and $\mathbf{E}(\mathbf{r})$ is the electric field at angular frequency $\omega$. 

In analogy to 3D-CWT, I-EIM assumes a separable ansatz $\mathbf{E}(\mathbf{r})=\mathbf{E_\text{2D}}(x,y)E_\text{1D}(z)$, which implicitly corresponds to a strictly TE-polarized field. Since this ansatz does not exactly satisfy Eq.~\ref{eq.11}, a residual field $\mathbf{R}(\mathbf{r})$ is introduced:
\[
\nabla \times \nabla \times \mathbf{E_\text{2D}}(x,y)E_\text{1D}(z) +\mathbf{R}(\mathbf{r})= \]\[\left(\frac{\omega}{c}\right)^2 \epsilon(\mathbf{r}) \mathbf{E_\text{2D}}(x,y)E_\text{1D}(z),\tag{9}\label{eq.12}\]
The method then seeks $\mathbf{E_\text{2D}}$, $E_\text{1D}$ and $\omega$ that minimize the residual norm squared $\iiint|\mathbf{R}|^2\rightarrow \text{min}$. The effective-index approach implies that $\mathbf{E_\text{2D}}(x,y)$ and $E_\text{1D}(z)$ satisfy the vector 2D and scalar 1D wave equations:
\[
\nabla \times \nabla \times \mathbf{E}_{2\mathrm{D}} =
\left(\frac{\omega}{c}\right)^2 \bar{\epsilon}_{2\mathrm{D}}(x,y)\,\mathbf{E}_{2\mathrm{D}},\tag{10}\label{eq.13}
\]
\[
-\,\frac{\partial^2 E_{1\mathrm{D}}}{\partial z^2}
+ \beta^2 E_{1\mathrm{D}} =
\left(\frac{\omega}{c}\right)^2 \bar{\epsilon}_{1\mathrm{D}}(z)\,E_{1\mathrm{D}},
\tag{11}\label{eq.14}\]
where $\bar{\epsilon}_{2\mathrm{D}}$ and $\bar{\epsilon}_{1\mathrm{D}}$ are yet unknown dielectric constant profiles of an effective periodic 2D photonic crystal and a 1D multilayer slab, respectively. 

As shown in Ref.~\cite{lang2025assessment}, inserting these equations into Eq.~\ref{eq.12}, neglecting the surface term $
\nabla \cdot \ \mathbf{E}_{2\mathrm{D}} $ and minimizing the residual norm yields the following relations:
\[\bar{\epsilon}_{1\mathrm{D}}(z)=\begin{cases}
\frac{\epsilon_\text{hole}\iint_{\text{hole}} \left| \mathbf{E}_{2\mathrm{D}} \right|^2  dxdy+\epsilon_\text{bulk}\iint_{\text{bulk}} \left| \mathbf{E}_{2\mathrm{D}} \right|^2  dxdy}{\iint_{\text{hole+bulk}} \left| \mathbf{E}_{2\mathrm{D}} \right|^2  dxdy}, & \text{if } z \in \text{PhC}, \\
\epsilon(z),  & \text{if } z \notin \text{PhC}
\end{cases}
\tag{12}\label{eq.15}\]
\[\bar{\epsilon}_{2\mathrm{D}}(x,y)=\]
\[\begin{cases} \frac{\epsilon_{\text{hole}}\int_{z\in\text{PhC}} \left| E_{1\mathrm{D}} \right|^2dz+\int_{z\notin\text{PhC}}\epsilon(z) \left| E_{1\mathrm{D}} \right|^2dz}{\int_{\text{all}}\epsilon(z) \left| E_{1\mathrm{D}} \right|^2dz}, & \text{if } (x,y) \in \text{hole}, \\ \frac{\epsilon_{\text{bulk}}\int_{z\in\text{PhC}} \left| E_{1\mathrm{D}} \right|^2dz+\int_{z\notin\text{PhC}}\epsilon(z) \left| E_{1\mathrm{D}} \right|^2dz}{\int_{\text{all}}\epsilon(z) \left| E_{1\mathrm{D}} \right|^2dz}, & \text{if } (x,y) \in \text{bulk} \end{cases} \tag{13}\label{eq.16}.\]
These expressions reveal that the effective 1D dielectric constant coincides with the original profile $\epsilon(z)$ outside the PhC layer, while inside the PhC it depends on $\epsilon(x,y)$ and the effective 2D solution $\mathbf{E_\text{2D}}(x,y)$. Conversely, the effective 2D dielectric constant depends on overlap integrals of $E_\text{1D}$ with the PhC hole and bulk regions, as well as other PCSEL layers. Thus, the 1D and 2D problems are mutually coupled through their respective field profiles.

Ref.~\cite{lang2025assessment} proposes a natural iterative method to find self-consistent solutions and effective index profiles, starting from an initial guess of $\bar{\epsilon}_{2\mathrm{D}}(x,y)$. With this guess, the 2D eigenproblem in Eq.~\ref{eq.12}
is solved using the open-source software package MPB\cite{johnson2001block}, and the mode of interest (A or B in our study) is selected. Its field profile $\mathbf{E_\text{2D}}(x,y)$ determines $\bar{\epsilon}_{1\mathrm{D}}(z)$ according to Eq.~\ref{eq.15}. The latter is inserted into Eq.~\ref{eq.14} together with the angular frequency $\omega$ of the preliminary 2D mode, thus leading to a vertical field profile $E_\text{1D}(z)$ and its associated effective index $n_{\text{eff}}$. Similarly to our realization of 3D-CWT, we use the FDE eigensolver with a sufficiently large computational domain in order to find out that vertical field profile. Alternatively, one could also use the semi-analytical transfer-matrix method for that purpose. Afterwards, that solution is used in Eq.~\ref{eq.16} to derive $\bar{\epsilon}_{2\mathrm{D}}(x,y)$. The iterative process repeats several cycles until both solutions converge and a certain convergence criterion is met. As such a criterion, we require the iterative step of the traced propagation constant to decrease below the step tolerance: $\Delta \beta/\beta<10^{-5}$. In practice, three to five iteration cycles were typically sufficient to meet this criterion in our study.

Now, it makes sense to highlight one important limitation of I-EIM, at least in its currently presented basic formulation. The method does not allow us to reliably estimate the radiation constants of the converging modes, characterized by real-valued eigenfrequencies. Indeed, the modeled effective system is infinitely periodic in the $xy$-plane, so that no in-plane losses may in principle occur. Additionally, the effective 1D field solution represents a fundamental guided mode, which is typically a bound state with evanescently decaying tails, meaning that it does not couple to the radiative continuum of outgoing waves. On the other hand, we know from theory that, for example, double-lattice PhC patterns lead to radiative symmetry-broken quasi-BIC states even if they are vertically well confined, while in the single-lattice case, at least symmetry-unprotected at-$\Gamma$ modes C and D are radiative. To rephrase, the effective 1D description underlying the I-EIM cannot adequately capture the radiation mechanism associated with distributed out-of-plane Bragg scattering at PhC slab features. 

Next, the resulting vertical field profile $E_{\text{1D}}(z)$ of the traced mode defines the normalized PhC confinement factor $\Gamma_{\text{PhC}}=\int_{\text{PhC}}\bar{\epsilon}_{1\mathrm{D}}(z)|E_{\text{1D}}(z)|^2 dz/\int_{\text{all}}\bar{\epsilon}_{1\mathrm{D}}(z)|E_{\text{1D}}(z)|^2 dz$, which is mode-specific. This constitutes a key difference from the 3D-CWT approximation, in which all four eigenmodes A, B, C, and D share a common vertical electric-field profile $E_{\text{1D}}(z)$ and PhC confinement factor $\Gamma_{\text{PhC}}$.  

In this study, we trace only BIC states A and B at the second-order $\Gamma$ point to directly compare their associated effective indices and PhC confinements with the results of 3D-CWT and rigorous methods. Once the traced resonant mode with $\omega_{i\in\{\text{A,B}\}}$ is obtained from the 2D eigensolver, the angular frequencies of the other three complementary modes $\omega_{j}$ follow from that solver as well. Therefore, the spatial eigenfrequency of each mode can be found as:
\[\overline{\nu}_{j\in\{\text{A,B,C,D}\}}=\frac{\omega_j}{\omega_i}\beta_i,\tag{14}\label{eq.17}\]
where $i$ is the index of the traced mode, so that the spatial frequency of that mode is equal to its propagation constant from the last iteration step (see Eq.~\ref{eq.14}): $\overline{\nu}_i=\beta_i$. The coupling coefficients $\kappa_{\text{1D}}, \kappa_{\text{2D}}$ are then extracted within the generalized  CWT framework, using the Eqs.~\ref{eq6} and \ref{eq7} for the Type I PCSEL, for which the A and B modes can be safely treated as non-radiative at sufficiently small hole diameters, as demonstrated below. For the Type II PCSEL, only the real-valued $\kappa_{\text{2D+}}$ can be evaluated via Eq.~\ref{eq3}, since the radiation constants of the four modes are not available within the I-EIM, rendering the information required to determine complex-valued coupling coefficients incomplete.

In contrast to the $\kappa_{\text{1D}}$, the two-dimensional coupling coefficients have to be subsequently normalized by the confinement factor $\Gamma_{\text{PhC}}$ of the traced mode. The rationale for this renormalization is provided in the Appendix. All coupling-coefficient values reported in this study are based on tracing mode A. We have preliminarily verified that coupling coefficients obtained by tracing mode B differ only marginally from those presented values, with relative deviations not exceeding $\sim2\%$ over all geometric parameter domains considered in the study. This indicates a good self-consistency of the I-EIM method. To conclude this section, we note that the effective indices of the traced modes A and B are evaluated as $n_{\text{eff},i}=\beta_ic/\omega_i$.
 
\subsection{\label{sec:level2}Finite element method}
The first rigorous method employed in this study is the full three-dimensional FEM solver implemented in COMSOL Multiphysics. This approach computes stationary electromagnetic fields as intrinsically three-dimensional entities, without imposing any enforced factorization of the form ${\mathbf{E}=E_\text{2D}(x,y)E_{\text{1D}}(z)}$.  Moreover, strict TE-polarization of the fundamental A, B, C, and D modes is not assumed within the FEM framework, allowing for a considerable out-of-plane electric-field component $E_z(x,y,z)$ component. 

Periodic boundary conditions are applied along the $x$- and $y$-axes, rendering the structure infinite in the lateral dimensions. The multilayer stack is terminated at the top and bottom by perfectly matched layers (PMLs). To minimize spurious mode overlap with the PMLs, a 200-nm air buffer layer is introduced above the ITO layer, whereas the lower PML is placed directly beneath the n-type AlGaN cladding layer, which has a thickness of 2000 nm. Our convergence tests have shown that a physics-controlled mesh with the element-size regime set to “finer” provides sufficient accuracy for determining the resonance eigenfrequencies and Q factors of the PCSEL structures under study.

Based on this method, radiation constants are evaluated as $\alpha_i=\beta_0/Q_i$, while the spatial frequencies are given by $\overline{\nu}=\omega_in_\text{eff}/c$, where $n_\text{eff}$ is approximated by the effective index of the 1D multilayer slab introduced in the 3D-CWT model. Note that this approximation is sufficient for determining the 1D and 2D coupling coefficients, as these are more sensitive to the frequency gaps between the modes than to their absolute values. 

\subsection{\label{sec:level2}Fourier modal method}
The second rigorous method employed in this study is the FMM. In this approach, the square unit cell of the PhC and of the entire PCSEL structure is periodically extended along the $x$- and $y$-directions. The computational domain is terminated 200 nm above the ITO layer and 500 nm below the cladding-substrate interface, thereby ensuring sufficient optical isolation of vertically confined resonant guided modes. The number of wavevectors of the Fourier modes was truncated at 120 with a circular domain of the reciprocal space, while the number of square mesh cells in $xy$-plane was set to $100\times100$. Optical excitation of guided resonances is implemented by means of the s- and p-polarized plane waves incident on the ITO layer. To excite the non-radiative at-$\Gamma$ BIC states of the single-lattice Type I PCSEL, the plane-wave incidence angle is slightly tilted to $\theta=0.01^\circ$. Although this condition does not correspond exactly to the second-order $\Gamma$ point, it enables a fairly accurate estimation of the demanded eigenfrequencies. By contrast, in the Type II PCSEL, all four modes are generally radiative; accordingly, excitation is performed using a normally incident plane wave  ($\theta=0^\circ$).

A parametric sweep of the excitation frequency yields reflection and transmission spectra of the structure, capturing both resonant dynamics associated with the embedded PhC grating and spectrally broader Fabry–Pérot-like longitudinal modes arising from multiple reflections at layer interfaces~\cite{song2018first}. Once the fundamental quasi-TE modes A, B, C, and D are identified in the transmission (or equivalently, reflection) spectrum, their linewidths and associated quality factors are extracted via Fano resonance fitting. An analogous Fano-resonance fitting model was applied, for example, in Ref.~\cite{lee2012observation} to experimentally measured reflectivity spectra of photonic crystal slabs. The corresponding radiation constants are then evaluated as $\alpha_i=\beta_0/Q_i$. The spatial frequencies are estimated from the angular frequencies, in analogy to the FEM approach. 

As will be shown with concrete data in the following section, at large hole filling factors ($ff>11\%$) in the Type I PCSEL, spectral overlap between the degenerate C and D modes and an additional broad resonance prevent accurate extraction of the corresponding Q factors and, consequently, the radiation constants $\alpha_{\text{C,D}}$. A similar mode-mixing issue arises for mode B at $ff=20\%$.

 \section{\label{sec:level1}RESULTS AND DISCUSSION}
\subsection{\label{sec:level2_a}Single-lattice photonic crystal}
\begin{figure*}
\includegraphics[width=0.5\textwidth]{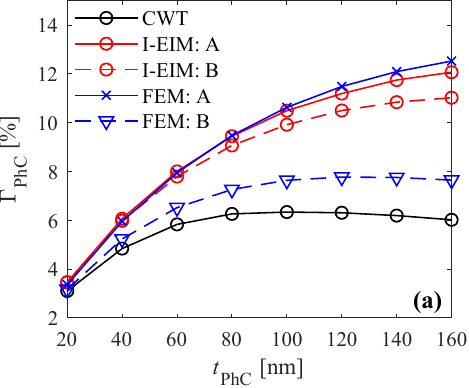}
\includegraphics[width=0.5\textwidth]{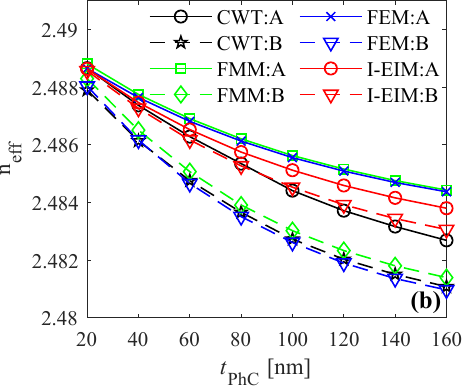}
\caption{\label{fig:SingLat_tPhCvar} (a) PhC confinement factors as functions of the PhC thickness, obtained by 3D-CWT, FEM, and I-EIM. $ff$ is fixed to 10 \%. (b) Effective indices as functions of $t_{\text{PhC}}$, obtained from $\beta_0c/\omega$ in the rigorous methods and from $\beta c/\omega$ in the effective-index-based methods. $ff$ is fixed to 10 \%.}
\end{figure*}

In this section, we focus on the highly symmetric Type I pattern, for which 3D-CWT\cite{liang2011three,liang2012three} predicts non-radiative A and B modes ($\alpha_{\text{A,B}}=0$) and degenerate C and D modes ($\overline{\nu}_C=\overline{\nu}_D, \alpha_{C}=\alpha_{D}$). Figures~\ref{fig:SingLat_ffvar}(a) and \ref{fig:SingLat_tPhCvar}(a) show that, by construction of 3D-CWT, the photonic-crystal confinement factor $\Gamma_{\text{PhC}}$ is identical for all four modes, whereas the other methods resolve mode-dependent electric-field profiles and, consequently, mode-dependent values of $\Gamma_{\text{PhC}}$. As evident from the plots, 3D-CWT systematically underestimates $\Gamma_{\text{PhC}}$ when compared with the mode-specific confinement factors $\Gamma_{\text{PhC,A}}$ and $\Gamma_{\text{PhC,B}}$ obtained from the other methods. 
Note that modes C and D exhibit even weaker confinement in the PhC layer (not shown here), making the CWT-based $\Gamma_\text{PhC}$ an adequate estimate of an average confinement factor for all four modes, rather than only for modes A and B. As evidenced in the corresponding plots, I-EIM agrees well with FEM in predicting $\Gamma_\text{PhC}$ for mode A, but yields significantly higher values for mode B. 
 
Next, we observe a good agreement between FEM and FMM with regard to the modal effective indices $n_\text{eff}$, as shown in Figs.~\ref{fig:SingLat_ffvar}(b) and \ref{fig:SingLat_tPhCvar}(b). 3D-CWT exhibits good agreement with FEM and FMM in predicting $n_\text{eff,B}$, but shows reduced agreement for $n_\text{eff,A}$. By contrast, I-EIM yields close agreement with the rigorous methods for the effective index of mode A, but shows noticeable deviations for mode B. The consistent agreement between FMM and FEM suggests an overall higher accuracy of these approximation-free methods. Note that both the mode splitting $n_\text{eff,A}-n_\text{eff,B}$ and the difference $\Gamma_\text{PhC,A}-\Gamma_\text{PhC,B}$ vanish, as expected, as $ff$ or $t_\text{PhC}$ approaches zero.
\begin{figure*}
\includegraphics[width=0.5\textwidth]{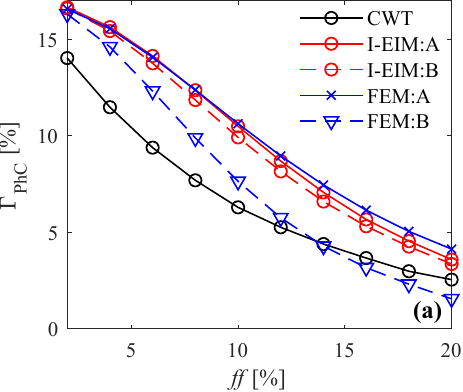}
\includegraphics[width=0.5\textwidth]{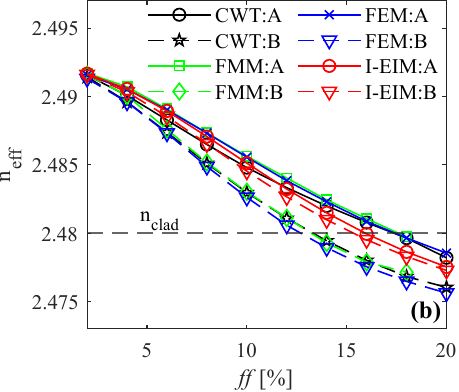}
\caption{\label{fig:SingLat_ffvar}(a) PhC confinement factors based on three methods as functions of the hole filling factor in the Type I PCSEL. 3D-CWT, unlike other methods, assumes a common vertical electric-field profile for all four modes, and thus $\Gamma_{\text{PhC}}$. $t_{\text{PhC}}$ is fixed to 100 nm. (b) Effective indices as functions of $ff$, obtained from $\beta_0c/\omega$ in the rigorous methods and from $\beta c/\omega$ in the effective-index-based methods. The photonic-crystal thickness $t_{\text{PhC}}$ is fixed to 100 nm. The horizontal dashed line indicates the refractive index of the n-type cladding, below which at-$\Gamma$ BIC states A and B transition into leaky quasi-BIC states.}
\end{figure*}

\begin{figure*}
\includegraphics[width=0.5\textwidth]{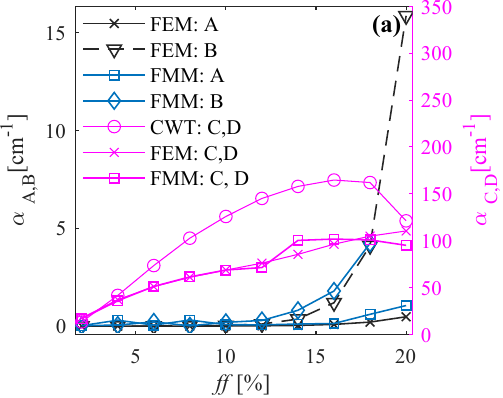}
\includegraphics[width=0.5\textwidth]{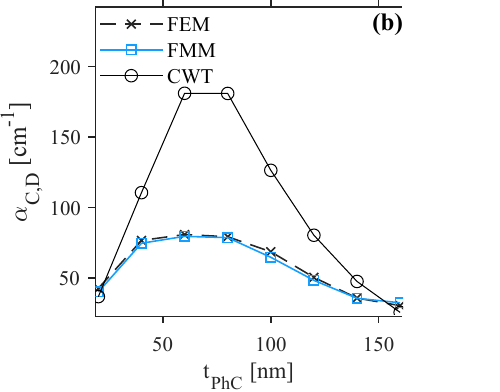}
\caption{\label{fig:SingLat_alpha} (a) Radiation constants of the fundamental A,B,C and D modes, plotted in different scales (left and right $y$-axes) as functions of the $ff$. $\alpha_{\text{B}}$ is not plotted at $ff=20\%$ due to a strong spectral overlap of mode B with another excited resonance. The PhC thickness is set to 100 nm. (b) Radiation constants of modes C and D as functions of PhC thickness at fixed $ff=10 \%$. The radiation constants of modes A and B are omitted from the plot due to their negligible values  ($\ll$1 $\text{cm}^{-1}$) .}
\end{figure*}

Figures~\ref{fig:SingLat_alpha}(a,b) show the modal radiation constants $\alpha_i$ as functions of the hole filling factor $ff$ and the PhC thickness $t_{\text{PhC}}$.  Overall, we observe very good agreement between FEM- and FMM-based data, while 3D-CWT generally predicts higher losses for states C and D and lossless behavior for BIC states A and B, regardless of the hole diameter. As shown in Fig.~\ref{fig:SingLat_alpha}(a), the accuracy of FMM slightly deteriorates when the degenerate C and D modes spectrally overlap with an additional resonance at $ff>$ 11\% for $t_{\text{PhC}}=100$ nm. The same mode-mixing problem occurs for mode B at $ff=20\%$, rendering $\alpha_{\text{B}}$ virtually unidentifiable at that point.

In contrast to the lossless BIC states predicted by 3D-CWT, the rigorous methods exhibit an exponential increase in $\alpha_{\text{B}}$ and, with some delay, of $\alpha_{\text{A}}$ at larger hole diameters. It is evident that at $ff\gtrsim12\%$ the elevated values of $\alpha_{\text{B}}$ become difficult to ignore. In the following, we provide a qualitative explanation of this behavior, focusing on mode B for concreteness, while the same arguments apply to mode A at larger hole diameters. 

For $ff\lesssim13\%$, the effective index of mode B remains above the refractive index of the n-type cladding, as shown in Fig.~\label{fig:SingLat_ffvar}(b). In this range, the guided resonance remains vertically confined, and the theory of at-$\Gamma$ BIC states predicts a perfectly lossless mode B. The finite Q factors obtained from rigorous methods arise primarily from finite mesh discretization, which disturbs both the continuous rotational symmetry of the circular hole and the discrete $C_4$ symmetry of the underlying square lattice. In principle, mesh refinement should and does lead to arbitrarily high Q factors, albeit at the cost of increased computational effort. In FMM, the discretization is limited to the $xy$-plane; however, this method introduces additional limitations, such as truncation in $k$-space and limited resolution of reflection and transmission spectra. 

Next, as the effective index of the BIC state approaches that of the n-type cladding from above, the mode begins to leak through that “potential barrier” into the GaN substrate, analogous to quantum tunneling. It becomes increasingly difficult to suppress this power leakage through further mesh refinement or enlargement of the computational domain. The theory of leaky modes in waveguides \cite{snyder1983optical} describes the asymptotic form of such fields as proportional to $\exp(ikz\sqrt{n_{\text{clad}}^2-n_{\text{eff}}^2})$ as $z\rightarrow\infty$, indicating a transition from an evanescently decaying field to an oscillatory plane wave at $n_{\text{eff}}=n_{\text{clad}}$. A qualitatively similar behavior is observed in our PCSEL system, although it is clear that the transition to leaky modes does not manifest itself as an abrupt, sharp increase in the FEM- and FMM-based values of $\alpha_\text{B}$ or $\alpha_\text{A}$. Nevertheless, the theory of leaky modes suggests that further refinement of rigorous numerical models should, in principle, lead to a sharper transition and a steeper slope of the radiation constants. Note that symmetry protection of BIC states A and B does not prevent them from becoming leaky at their respective critical hole diameters. 

\begin{figure}
\includegraphics{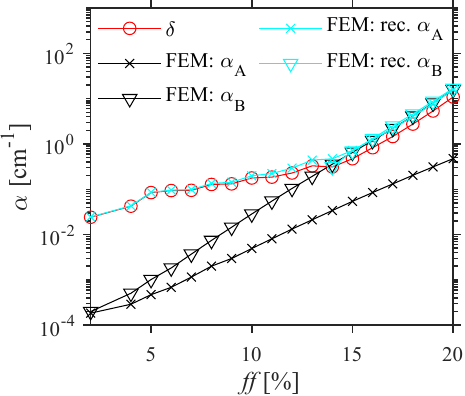}
\caption{\label{fig:SingLat_delta} Radiation constants $\alpha_{\text{A,B}}$ derived from FEM (black curves) and their CWT-compliant counterparts (cyan curves) as functions of $ff$. The error $\delta$ is shown in red. Data points of the reconstructed $\alpha_\text{B}$ are omitted from the semi-logarithmic plot for $ff<14\%$, as $\alpha_\text{B,rec.}$ takes slightly negative values (down to approximately $-0.1$  $\text{cm}^{-1}$) in this range.}
\end{figure}

In the following, we show that, unlike rigorous methods, the hybridized CWT framework captures this anticipated sharp phase transition from theoretically lossless BIC states to their leaky quasi-BIC extensions. To do so, we first introduce the following absolute error function, which quantifies the discrepancy between a rigorous method and the associated CWT framework:
\[\delta=\sqrt{\frac{|\alpha_{\text{A}}-\alpha_{\text{A,rec}}|^2+|\alpha_{\text{B}}-\alpha_{\text{B,rec}}|^2}{2}},\tag{15}\label{eq.8}\]
where values $\alpha_{\text{A,B; rec}}$ denote best-fit parameters, reconstructed from the CWT framework using either FEM or FMM data as input. Next, we focus on FEM-based radiation constants, which exhibit a slightly more analytical behavior as functions of $ff$, and plot $\delta$ along with $\alpha_{\text{A,B}}$ and their CWT-compliant counterparts $\alpha_{\text{A,B; rec}}$ in Fig.~\ref{fig:SingLat_delta}. The logarithmic scale enhances visibility over a wide dynamic range. The results clearly show that both the error metric $\delta$ and the CWT-compliant radiation constants undergo an exponential increase starting at $ff\approx14 \%$. Note that the discrepancy $\delta$ between the CWT framework and the corresponding rigorous method is gradually increasing with $ff$ even at small hole diameters, which likely reflects the perturbative treatment\cite{persson2025finite} of the scattering hole features within the 3D-CWT theory.

Overall, while the idealized 3D-CWT predicts perfect symmetry protection ($\alpha_{A,B}=0$) and rigorous methods reveal exponentially growing radiation constants accompanied by an increasing asymmetry $\alpha_\text{B}-\alpha_\text{A}$, the hybridized CWT framework approximately preserves symmetry protection ($|\alpha_\text{A,B;rec}|<1$ $\text{cm}^{-1}$) only below $ff\approx14 \%$. For larger hole diameters, it instead predicts exponentially increasing radiative rates with nearly balanced levels ($\alpha_\text{A,rec}\approx\alpha_\text{B,rec}$). That critical $ff$ value of $14\%$ closely matches the anticipated phase transition of mode B into a leaky mode at the condition $n_\text{eff}=n_{\text{clad}}$, corresponding to $ff\approx13\%$. It is reasonable to assume that the delayed leakage of mode A at a larger hole filling factor shifts the joint critical point $ff\approx14\%$ slightly above the leakage threshold of mode B. Notably, even at large hole diameters, the CWT framework maintains an approximate balance between the radiation constants of modes A and B, suggesting that a weakened form of symmetry protection remains implicitly enforced even in the leaky-mode regime. On the other hand, our results from rigorous methods do not support this balanced relation between the two modes, suggesting that a more nuanced understanding beyond the CWT framework is required in the leaky-mode regime.

In summary,  the hybridized CWT framework provides a balanced description bridging the idealized 3D-CWT and more realistic rigorous approaches, capturing one joint phase transition point of the two at-$\Gamma$ BIC states in accordance with the theory of leaky modes. The error metric $\delta$ provides a convenient indicator of this critical point. 

\begin{figure*}
\includegraphics[width=0.5\textwidth]{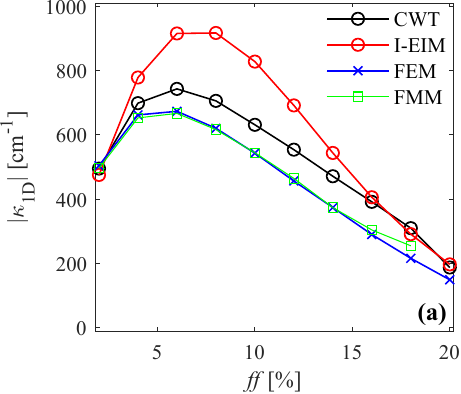}
\includegraphics[width=0.5\textwidth]{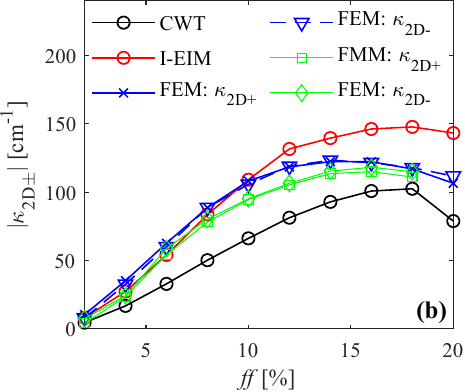}
\caption{\label{fig:SingLat_ffvar_kappa} (a) Absolute values of the one-dimensional coupling coefficient as a function of $ff$, based on the corresponding four methods. PhC thickness is set to 100 nm. The data point of FMM is omitted at $ff=20\%$ due to a missing input value of $\alpha_{\text{B}}$. (b) Absolute values of 2D coupling coefficients as functions of $ff$, showing the expected equality $\kappa_{\text{2D+}}\approx\kappa_{\text{2D-}}$ The data point of FMM at $ff=20\%$ is omitted for the same reason as stated above. PhC thickness is fixed to 100 nm. }\end{figure*}

In the following step, we evaluate coupling coefficients as described previously by feeding $\alpha_{\text{C,D}}$ and $\overline{\nu}_{\text{A,B,C,D}}$  into Eqs.~(\ref{eq1},\ref{eq2}) and fitting the $\alpha_{\text{A,B}}$ and coupling coefficients within the CWT framework. In Figs.~\ref{fig:SingLat_ffvar_kappa}(a)  and \ref{fig:SingLat_tPhCvar}(a) the parameter $\kappa_{\text{1D}}$ obtained from FEM and FMM shows close agreement, at least for $ff<16\%$. The increasing discrepancy observed at larger hole diameters can be attributed to the previously discussed unintended mode-mixing of modes C and D with another excited resonance, which makes a more precise extraction of $\alpha_{\text{C,D}}$ challenging in FMM. 

Figure~\ref{fig:SingLat_ffvar_kappa}(b) further demonstrates a relatively good agreement between FEM and FMM for $\kappa_{\text{2D}}$. Overall, 3D-CWT provides a more accurate and consistent match to FEM and FMM results for $\kappa_{\text{1D}}$, whereas the I-EIM approach performs better in predicting $\kappa_{\text{2D}+}$. Nevertheless, for $ff>10 \%$, a growing deviation between I-EIM and the rigorous methods in $\kappa_{\text{2D}+}$ becomes apparent. This deviation at large hole diameters reflects the growing discrepancy between the effective indices predicted by I-EIM and those obtained from rigorous methods, particularly for mode B. 

\begin{figure*}
\includegraphics[width=0.5\textwidth]{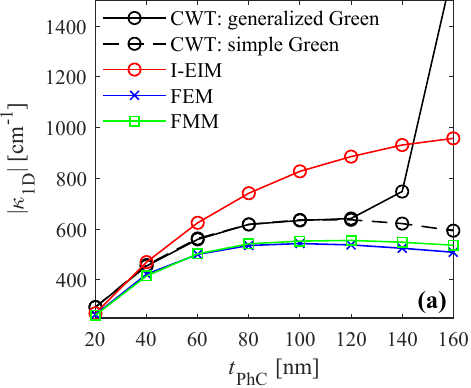}
\includegraphics[width=0.5\textwidth]{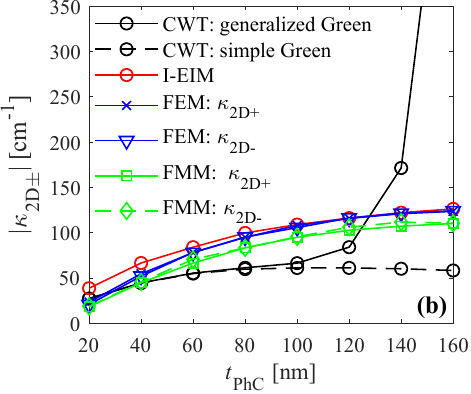}
\caption{\label{fig:SingLat_tPhCvar_kappa} Absolute values of the (a) one- and (b) two-dimensional coupling coefficients, plotted as functions of PhC thickness at $ff=10\%$. In 3D-CWT, a simplified exponential and, alternatively, a generalized Green function have been incorporated. The latter led to a clear divergence of both $\kappa$ at $t_{\text{PhC}}\gtrsim100$ nm.}
\end{figure*}

Finally, in 3D-CWT, we incorporate either simplified Green functions of the form $G(z,z')\sim e^{-i\beta_z|z-z'|}/(2\beta_z)$ or generalized Green functions \cite{peng2011coupled}, accounting for secondary wave reflections at multilayer interfaces. Figs.~\ref{fig:SingLat_tPhCvar_kappa}(a,b) reveals that the latter ansatz leads to a divergence of coupling coefficients $\kappa_{\text{1D,2D}}$ as $t_{\text{PhC}}$ exceeds $\sim$ 100 nm. We attribute this apparent divergence to a singularity of the Green functions themselves, which is a well-known issue in stratified lossless media \cite{gay2000accurate}. Such divergent behavior is not observed when using simplified Green functions that neglect secondary reflections at multilayer interfaces. Importantly, both Green-function formulations yield comparable results as long as $t_{\text{PhC}}$ does not exceed $\sim$100 nm. However, it is also evident that $\kappa_{\text{2D}}$, computed via the generalized Green functions, begins to diverge already at $\sim$~80 nm, approximately corresponding to the first-order Fabry–Pérot resonance of the PhC slab. 

 \subsection{\label{sec:level2_b}Double-lattice photonic crystal}
Here we consider the Type II PCSEL, wherein the double-lattice pattern lowers the $C_4$ symmetry of the square lattice, thus breaking the symmetry protection and forcing quasi-BIC modes A and B to irradiate at the second-order $\Gamma$ point. In addition, the degeneracy between modes C and D gets lifted according to the CWT framework (see Eqs.~(\ref{eq1},\ref{eq2})). All these changes are reflected in complex-valued coefficients $\kappa_{\text{1D}}$ and $\kappa_{\text{2D-}}$ and breaking of the symmetry condition $\kappa_{\text{2D-}}=\kappa_{\text{2D+}}$. In our simulated Type II system, we fix the filling factor $ff_2$ of the second hole to 4\%, while tuning only the filling factor $ff_1$ of the first hole and the lateral distance $\Delta$ between holes, illustrated in Fig.~\ref{fig:lattices}(b).

\begin{figure*}
\includegraphics[width=0.5\textwidth]{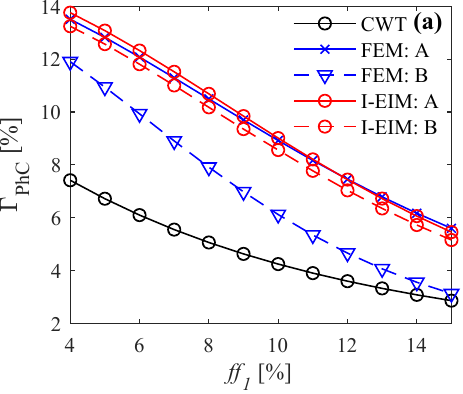}
\includegraphics[width=0.5\textwidth]{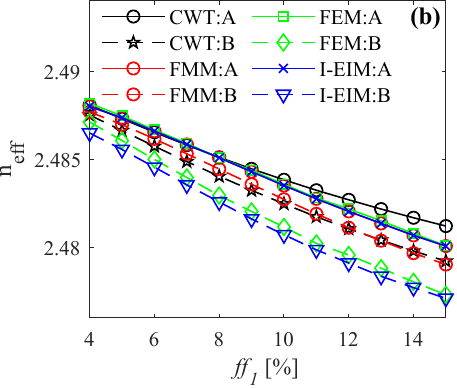}
\caption{\label{fig:DoubLat_ff1var_Gamma}(a) PhC confinement factors and (b) effective indices as functions of $ff_1$ of the first hole, derived from corresponding methods. $ff_2$ of the second hole is fixed to 4 \%, while the lateral distance between holes is $\Delta=0.46a$. }
\end{figure*}

\begin{figure*}
\includegraphics[width=0.5\textwidth]{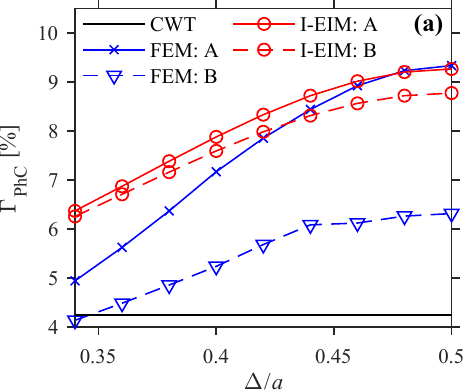}
\includegraphics[width=0.5\textwidth]{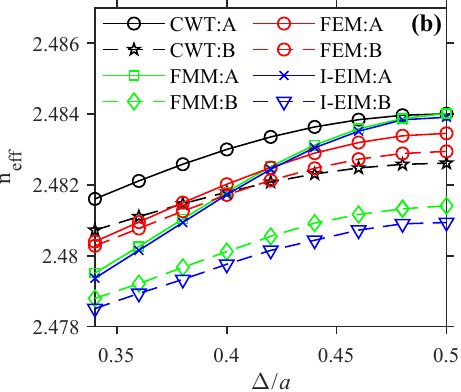}
\caption{\label{fig:DoubLat_dvar_Gamma}(a) PhC confinement factors and (b) effective indices as functions of the lateral distance between air holes. $ff_1$ and $ff_2$ are set to 10 \% and 4 \%, respectively.}
\end{figure*}

First, Figs.~\ref{fig:DoubLat_ff1var_Gamma}(b) and \ref{fig:DoubLat_dvar_Gamma}(b) reveal a great agreement in $n_{\text{eff}}$ between FEM and FMM. I-EIM agrees well in $\Gamma_{\text{PhC}}$ and $n_{\text{eff}}$ with rigorous methods only for mode A, provided that $\Delta$ is not too small. Similarly to the case of Type I PCSEL, 3D-CWT tends to underestimate $\Gamma_{\text{PhC}}$ values in comparison to other methods. In general, the resonance contrast $n_{\text{eff,A}}-n_{\text{eff,B}}$ predicted by FEM and FMM remains unmatched by both effective-index-based methods, while the large FEM-based contrast $\Gamma_\text{PhC,A}-\Gamma_\text{PhC,B}$ is not reproduced by I-EIM.
 
\begin{figure}
\includegraphics[width=0.5\textwidth]{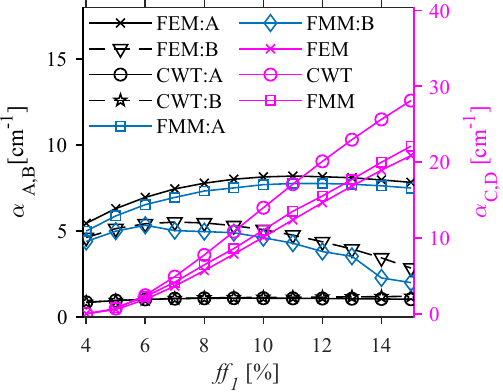}
\caption{\label{fig:DoubLat_alfa_ff1}Radiation constants of modes A, B (left $y$-axis) and C, D (right $y$-axis), plotted as functions of $ff_1$ based on three methods. The average of $\alpha_{\text{C}}$ and $\alpha_{\text{D}}$ is plotted in magenta, as the difference between them remains relatively small (<1$\%$). At $ff_1=4\%$, the values of $\alpha_{\text{C,D}}$ are extrapolated to 0  in the FMM, since none of the resonances can be excited at normal wave incidence. $\Delta$ is fixed to $0.46a$, while $ff_2$ is set to $4\%$.}
\end{figure}

\begin{figure*}
\includegraphics[width=0.5\textwidth]{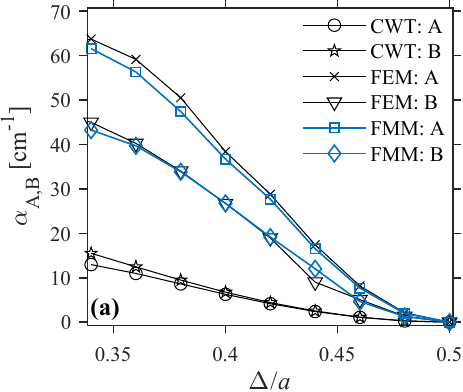}
\includegraphics[width=0.5\textwidth]{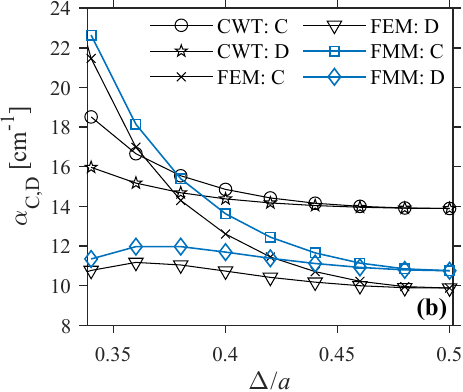}
\caption{\label{fig:DoubLat_alfa} Radiation constants of modes A, B (a) and C, D (b), plotted as functions of $\Delta/a$ based on three methods. $ff_1$ and $ff_2$ are fixed to $10\%$ and $4\%$, respectively, in both plots. At $\Delta=0.5a$, the values of $\alpha_{\text{A,B}}$ are extrapolated to 0  in the FMM, since none of the resonances can be excited at normal wave incidence.}
\end{figure*}

Next, in Figs.~\ref{fig:DoubLat_alfa_ff1} and \ref{fig:DoubLat_alfa}(a,b), we plot the radiation constants of the four modes. Overall, very good agreement between very good agreement between FEM and FMM is observed for all four $\alpha$'s. At the same time, it is notable that $\alpha_{\text{C,D}}$ obtained from the rigorous methods agree much better with their CWT-based counterparts than $\alpha_{\text{A,B}}$, which are significantly larger in the rigorous approaches. Beyond this quantitative discrepancy, a key difference between the CWT and rigorous methods lies in the asymmetry relation $\alpha_\text{B}>\alpha_\text{A}$, which holds in the former case but switches to $\alpha_\text{A}>\alpha_\text{B}$ in FEM and FMM cases.

\begin{figure}
\includegraphics[width=0.5\textwidth]{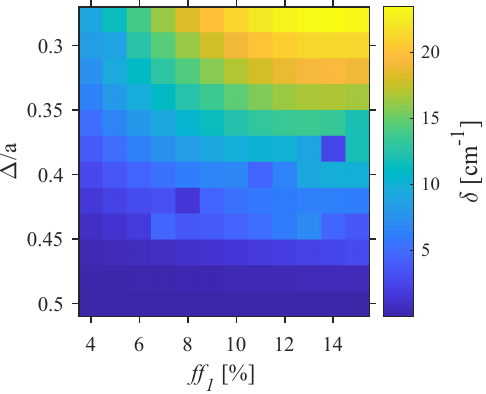}
\caption{\label{fig:DoubLat_delta} The absolute deviation $\delta$ between FEM-based and CWT-compliant radiation constants of modes A and B, plotted as a colormap function of $\Delta$ and $ff_1$.}
\end{figure}

\begin{figure*}
\includegraphics[width=0.5\textwidth]{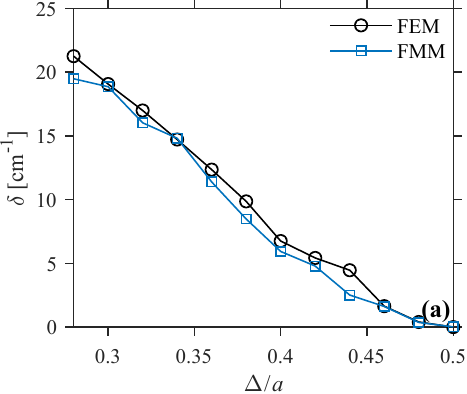}
\includegraphics[width=0.5\textwidth]{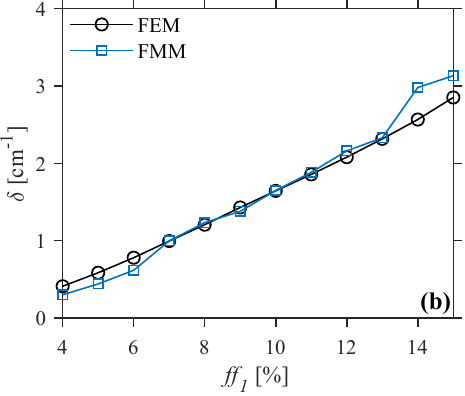}
\caption{\label{fig:DoubLat_delta2}The absolute deviation $\delta$ as a function of (a) $\Delta$ and (b) $ff_1$, obtained from FEM and FMM data. In (a), $ff_1$ is fixed at $10\%$, and in (b), $\Delta$ is fixed at $0.46a$. }
\end{figure*}

In analogy with Type I PCSEL, we plot $\delta$ from Eq.~\ref{eq.8} as a function of $ff_1$ and $\Delta$, as shown in Fig.~\ref{fig:DoubLat_delta}. For comparison, FMM-based values of $\delta$ are also provided along the cross-sections $ff_1=10\%$ and $\Delta=0.46a$ in Figs.~\ref{fig:DoubLat_delta2}(a) and (b), respectively. As seen, the FEM- and FMM-based values of $\delta$ are generally in good agreement. However, FEM-based data in Fig.~\ref{fig:DoubLat_delta}  also reveal abrupt variations in $\delta$, for example at $\Delta=0.44a$ and $7\%\leq ff_1\leq13\%$, as well as in other localized regions (e.g., dark spots in the colormap) of the illustrated domain. Our preliminary data show that these anomalies are not reproduced by the FMM results, suggesting that they arise from numerical artifacts specific to the FEM implementation.

The colormap plot shows a gradual increase of error $\delta$ with $ff_1$ and its decrease with $\Delta$. While in the Type I PCSEL, we observed a triggered at $ff\approx14 \%$ exponential increase of $\delta$ mainly due to growing $\Delta\alpha_\text{A}$, here instead a permanent near-linear increase takes place, arising from a balanced contribution of $\Delta\alpha_{\text{A}}$ and $\Delta\alpha_{\text{B}}$. Since the $C_4$ rotational symmetry of a square lattice is broken by the double-lattice pattern, the symmetry protection of the at-$\Gamma$ BIC states gets lifted, as evidenced by significant values of $\alpha_{\text{A,B}}$ in all methods, except when $\Delta=0.5a$. The resulting radiative states can therefore be classified as symmetry-broken quasi-BIC states, in contrast to the leaky quasi-BIC states demonstrated in Type I PCSEL. Note that in the symmetry-broken case, both modes A and B become radiative simultaneously as $\Delta$ deviates from $0.5a$, while in the Type I lattice with a single hole, mode B becomes leaky first.

\begin{figure*}
\includegraphics[width=0.5\textwidth]{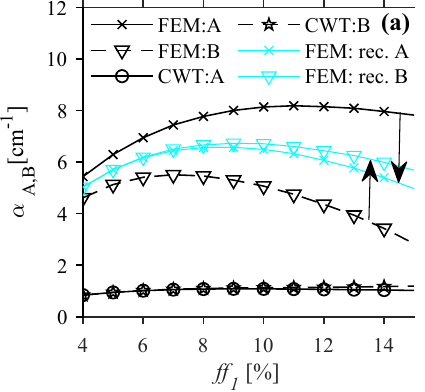}
\includegraphics[width=0.5\textwidth]{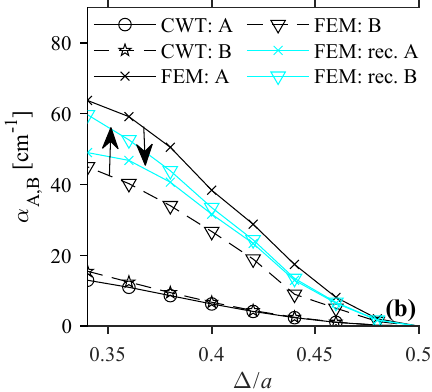}
\caption{\label{fig:AlfaABrecon} Radiation constants of modes A and B as functions of (a) $ff_1$ and (b) $\Delta$, derived from FEM and 3D-CWT as well as reconstructed from CWT framework combined with FEM-based data.}
\end{figure*}

On another note, a crucial difference becomes evident between the asymmetry ($\alpha_\text{B}>\alpha_\text{A}$) predicted by 3D-CWT and the CWT framework (see Figs.~\ref{fig:AlfaABrecon}(a,b) for the FEM-based fits) and the opposite condition  ($\alpha_\text{A}>\alpha_\text{B}$) obtained from the rigorous methods. In this sense, the error $\delta$ quantifies both the magnitude of this asymmetry between the quasi-BIC modes and the discrepancy between the CWT framework and a chosen rigorous method. Assuming that the asymmetry sign predicted by the rigorous methods is physically correct, this indicates a fundamental limitation of the CWT framework, for which no obvious remedy exists to the best of our knowledge.  

Next, we analyze the function $\delta(ff_1,\Delta)$ in Fig.\ref{fig:DoubLat_delta}(a) in more detail. Note that $\delta$ becomes relatively small when the lateral hole distance $\Delta$ is above $0.44a$, or the first hole filling factor becomes comparable to the second one $ff_1\sim ff_2$. Otherwise, the two-dimensional parameter is gradually increasing along $ff_1$ and $\Delta$. For example, $\delta$ crosses the 10 $\text{cm}^{-1}$ value at $ff_1=6 \%$ and $\Delta=0.30a$, but for $\Delta=0.36a$ the threshold shifts to $ff_1=10\%$. Between these two points, the total hole filling factor $ff_1+ff_2$ spans the range of $10-14\%$. For $\Delta>0.44a$, the value of $\delta$ remains below 5 $\text{cm}^{-1}$ at least in our range of interest $ff_1\leq15\%$, thus indicating a relatively weak asymmetry between modes A and B. Air holes with $ff_1=15\%$ and $ff_2=4 \%$ come into contact at $\Delta\approx0.23a$, implying that an additional bulk buffer of roughly the average hole diameter $(d_1+d_2)/2$ is required to reach $\Delta\approx0.46a$ and thus to keep $\delta$ small.
 If the hole diameters are comparable ($ff_1\approx ff_2=4\%$), then the threshold of 5 $\text{cm}^{-1}$ is respected only for $\Delta>0.36a$. The holes of that small diameter would come into contact at $\Delta\approx0.16a$, implying that an additional distance of $0.2a$, which is roughly equal to the hole diameter, is necessary to keep $\delta$ relatively small. In other words, as a qualitative rule, we propose that large hole diameters and bulk buffer gaps smaller than the average hole diameter lead to pronounced mode asymmetry and a substantial discrepancy between the rigorous methods and the CWT framework. As in the Type I PCSEL, we qualitatively observe feature-size-dependent accuracy of the CWT framework, which is likely associated with the perturbative nature of the 3D-CWT theory \cite{persson2025finite}. The only difference is that, in this case, the bulk gap between neighboring holes becomes an additional critical parameter.   

\begin{figure*}
\includegraphics[width=0.5\textwidth]{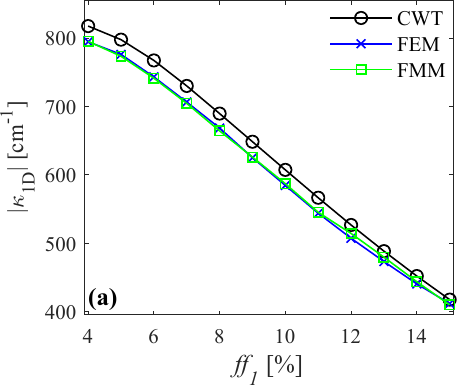}
\includegraphics[width=0.5\textwidth]{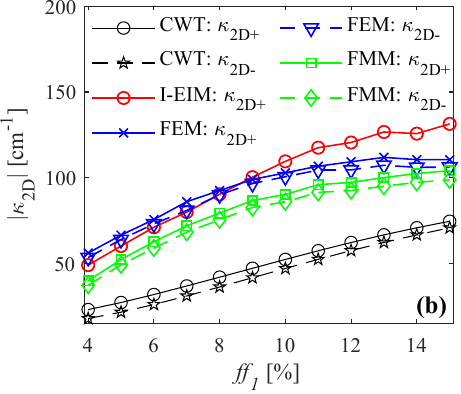}
\caption{\label{fig:DoubLat_ff1var_kappa} Absolute values of (a) one- and (b) two-dimensional coupling coefficients, plotted as functions of $ff_1$ based on different methods. $\Delta$ and $ff_2$ are fixed to $0.46a$ and $4\%$, respectively. In plot (a), only 3D-CWT, FEM, and FMM are represented, while in plot (b), additionally, the I-EIM-based $\kappa_{\text{2D+}}$ is plotted. }
\end{figure*}

\begin{figure*}
\includegraphics[width=0.5\textwidth]{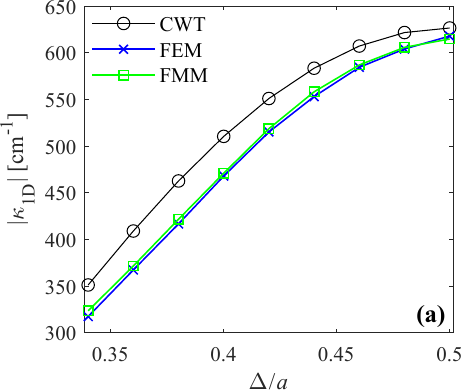}
\includegraphics[width=0.5\textwidth]{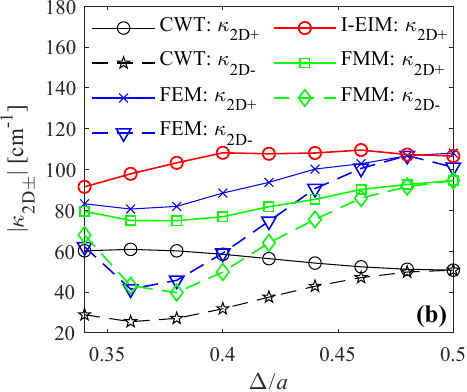}
\caption{\label{fig:DoubLat_dvar_kappa}  Absolute values of (a) one- and (b) two-dimensional coupling coefficients, plotted as functions of $\Delta/a$ based on different methods. $ff_1$ and $ff_2$ are fixed to $10\%$ and $4\%$, respectively.  I-EIM, lacking any data about radiation constants, is eligible only for defining $\kappa_{\text{2D+}}$ in plot (b). }\end{figure*}

Next, Figs.~\ref{fig:DoubLat_ff1var_kappa}(a,b) and \ref{fig:DoubLat_dvar_kappa}(a,b) demonstrate 1D and 2D coupling coefficients plotted as functions of $ff_1$ and $\Delta$, respectively. Again, we recognize a very good agreement in $\kappa_{\text{1D}}$ between FEM and FMM, while 3D-CWT delivers a slightly elevated estimation. As already mentioned earlier, I-EIM does not allow us to extract complex-valued $\kappa_{\text{1D}}$ and $\kappa_{\text{2D-}}$ due to missing radiation constants. Concerning $\kappa_{\text{2D}\pm}$, the best match is observed, as before, between FEM and FMM. Here, an inaccuracy of 3D-CWT becomes more evident as it predicts significantly lower absolute values $|\kappa_{\text{2D}\pm}|$. In contrast, I-EIM generates a more reliable estimation of $\kappa_{\text{2D+}}$, as it to some extent matches the rigorous methods.  Specifically, I-EIM-based $\kappa_{\text{2D+}}$ deviates from its counterparts obtained using rigorous methods when  $ff_1$ exceeds $\sim10\%$ at $\Delta=0.46a$ (see Fig.~\ref{fig:DoubLat_ff1var_kappa}(b)) or when $\Delta$ is reduced below $\sim0.44a$ at $ff_1=10 \%$ (Fig~\ref{fig:DoubLat_dvar_kappa}(b)). We assume that this disparity may reflect the gradual breakdown of the CWT framework, as discussed above. However, the exact relative position of I-EIM within the CWT framework cannot be fully assessed due to the absence of radiation constants and the associated complex-valued coupling coefficients for the Type II PCSEL. 
 
Finally, the pronounced downward bending of FMM- and FEM-based curves $|\kappa_{\text{2D}\pm}|$ in Fig.~\ref{fig:DoubLat_dvar_kappa}(b), as $\Delta$ decreases from $0.5a$, is likely another indication of the growing disparity between predictions of the rigorous methods and the CWT framework. Note that coupling coefficients derived from the FEM and FMM eigenfrequency inputs are by definition CWT-compliant. Therefore, the asymmetry $|\kappa_{\text{2D}+}|>|\kappa_{\text{2D}-}|$ (in analogy with $\alpha_\text{B}>\alpha_\text{A}$) remains intact within both 3D-CWT and the hybridized CWT framework. Assuming that the asymmetry $\alpha_\text{A}>\alpha_\text{B}$ predicted by the rigorous methods is physically correct, we infer that a proper formulation of the CWT framework, if achievable, should reproduce this asymmetry in $\alpha$ and enforce the corresponding relation $|\kappa_{\text{2D}-}|>|\kappa_{\text{2D}+}|$. In the current formulation, however, the asymmetry signs in $\alpha_\text{A,B}$ and $\kappa_{\text{2D}\pm}$ predicted by the hybridized CWT framework and 3D-CWT are likely unphysical, as they contradict the mutually consistent results of the rigorous methods. 

\section{\label{sec:level1}CONCLUSION}

In conclusion, we present a comparative analysis of rigorous and effective-index-based methods, along with a CWT framework relevant to square-lattice PhC patterns. We introduced an absolute deviation metric $\delta$ that, in the context of at-$\Gamma$ BIC states, quantifies the discrepancy between a given rigorous method and its corresponding hybridized CWT framework. 

For the single-lattice PCSEL, the hybridized CWT framework predicts a sharp, simultaneous transition of the at-$\Gamma$ BIC states into their leaky quasi-BIC counterparts at a critical hole filling factor $ff\approx 14\%$, manifested by an exponential increase in $\delta$. In contrast, 3D-CWT predicts unrealistically strong symmetry protection of the states in the leaky-mode regime, whereas rigorous methods indicate a thresholdless increase in the radiation constants and a growing asymmetry $\alpha_\text{A}-\alpha_\text{B}$ with increasing $ff$. Nevertheless, we expect that further refinement of the rigorous numerical models may better resolve and sharpen the individual phase transitions of modes A and B, thereby overcoming the inability of the CWT framework to distinguish between the states in the leaky-mode regime. 

In the double-lattice PCSEL, symmetry protection of the BIC states is lifted as a result of symmetry breaking of the underlying square lattice. However, a non-zero deviation metric $\delta$ reveals that both the hybridized CWT framework and 3D-CWT predict the mode asymmetry $\alpha_\text{B}>\alpha_\text{A}$, which is opposite in sign to that obtained from rigorous methods. Accordingly, the near-linear increase of $\delta$ as a function of $ff_1$ and $\Delta$ indicates a simultaneous increase in the asymmetry magnitude and in the discrepancy between the hybridized and rigorous approaches.  Given the consistent agreement between FEM and FMM throughout this study, we conclude that the CWT-based prediction of a lossier mode B is likely unphysical. Moreover, a proper formulation of the CWT framework, if achievable, is expected to yield the asymmetry $|\kappa_{\text{2D}-}|>|\kappa_{\text{2D}+}|$, which is opposite to the relation currently enforced by the CWT framework. It is worth emphasizing the distinction between symmetry-broken quasi-BIC states in Type I PCSELs and leaky quasi-BIC states in Type II PCSELs. 

Furthermore, as a qualitative rule, we found that $\delta$ becomes significant when either one of the hole diameters is sufficiently large, or the bulk buffer gap between adjacent holes becomes comparable to the average hole diameter. Thus, the validity limits of the CWT framework for double-lattice PCSELs are qualitatively identified in terms of the largest hole diameters and the smallest bulk geometric features. In the single-lattice PCSEL, the accuracy of the CWT framework is governed by the single hole diameter, but can be made arbitrarily high for well-confined modes ($ ff\ll14\%$), since $\alpha_{\text{A,B}}$ obtained from rigorous methods can, in principle, reach arbitrarily small values, as discussed in the main text. In the leaky-mode regime, however, the accuracy of CWT inevitably drops with $ff$ due to a growing discrepancy between $\alpha_\text{A}$ and $\alpha_\text{A, rec}$. 

In addition, we included the iterated effective-index method in our analysis. Although its position relative to other methods cannot be fully assessed due to the absence of a reliable estimate of the radiation constants, we nevertheless observed a partial lopsided agreement with the FEM and FMM results. In particular, I-EIM provided overall more accurate predictions of the real-valued $\kappa_{\text{2D}\pm}$, whereas 3D-CWT shows better agreement with rigorous methods for $\kappa_{\text{1D}}$. However, the relatively large contrast $\Gamma_{\text{PhC,A}}-\Gamma_{\text{PhC,B}}$ and the frequency splitting proportional to $n_\text{eff,A}-n_\text{eff,B}$, as predicted by FEM and FMM, could not be consistently reproduced by either I-EIM or 3D-CWT. 

Finally, the synergistic combination of the CWT framework with rigorous methods reveals the principal limits of validity of coupled-wave theory, particularly in predicting the subtle behavior of at-$\Gamma$ BIC states. At the same time, the hybridized CWT framework correctly captures symmetry-broken quasi-BIC states as well as the qualitative phase transition into leaky quasi-BIC states, in accordance with leaky-mode theory. In this sense, the mutually consistent rigorous methods provide a reliable benchmark and input for the CWT framework.  The proposed comparative technique can be readily extended to a broader class of PCSEL architectures and photonic-crystal patterns, thereby enabling a deeper and more comprehensive understanding of these devices. 

\begin{acknowledgments}
The project was funded by the Deutsche Forschungsgemeinschaft (DFG, German Research Foundation) under Germany’s Excellence Strategy – EXC-2123 QuantumFrontiers – 390837967. We acknowledge the financial support by the Volkswagen-Stiftung and the „Niedersächsisches Ministerium für Wissenschaft und Kultur“ within the Quantum Valley Lower Saxony Q1 project. Furthermore, the BMFTR funded project „QPIC“ within the Cluster4Future „QVLS iLabs“ supported this research. The authors thank Frederik Lüßmann, Daniel Stoll and Jana Hartmann for valuable support and discussions. 
\end{acknowledgments}

\section*{Conflict of Interest Statement }
The authors have no conflicts to disclose.

\section*{Data Availability Statement}

The data that support the findings of this study are available from the corresponding author upon reasonable request.

\appendix*
\section{Normalization of 2D coupling coefficients in I-EIM}
Here we argue that once $\kappa_{\text{2D}+}$ is derived from Eq.~\ref{eq7}  or Eq.~\ref{eq3} using I-EIM-based eigenfrequencies, it must be normalized by the confinement factor $\Gamma_{\text{PhC}}$. To motivate this requirement, we first consider the analytical CWT expression for $\kappa_{\text{2D}+}$ (for details, see e.g. \cite{inoue2022general,liang2011three}) :
\[
\kappa_{\text{2D}+}
=
\frac{k_0^2}{2 \beta_0 }
\sum_{\sqrt{m^2+n^2}>1}
\xi_{1-m,\,-n}\zeta^{(0,1)}_{y,m,n},\tag{A1}\label{eq.9}
\qquad
\]
where $k_0$ is the free-space wavenumber, $\beta_0=2\pi/a$, $\xi_{m,n}$ denote the spatial Fourier coefficients of the patterned PhC dielectric constant $\varepsilon_{\text{PhC}}
(x,y)=\varepsilon_{\text{avg}}+\sum_{m,n\neq0}\xi_{m,n}e^{-i\beta_0(mx+ny)}$ with $\varepsilon_{\text{avg}}$ representing the spatially averaged dielectric constant of the PhC layer.  According to \cite{liang2011three}, $\zeta^{(0,1)}_{y,m,n}$ takes the form:

\[\zeta^{(0,1)}_{y,m,n}\sim-k_0\iint\xi_{m,n-1}(z)G_{m,n}(z,z')E_{\text{1D}}(z)E_{\text{1D}}^*(z')dz'dz\]\[-\int\frac{\xi_{m,n-1}(z)|E_{\text{1D}}(z)|^2}{\varepsilon_\text{avg}(z)}dz\tag{A2}\label{eq.10}\]
where the integrals are evaluated over the entire multilayer structure, $E_{\text{1D}}(z)$ denotes the effective one-dimensional electric-field profile, and $G_{m,n}(z,z')$ are the Green’s functions defined in \cite{liang2011three}.  

Within the 3D-CWT method, the terms $\xi_{m,n}(z)$ are proportional to the material contrast $\Delta\varepsilon(z)$, and are therefore nonzero only inside the PhC region. Consequently, the integrals in the above expression receive contributions only from the PhC layer of finite thickness $t_\text{PhC}$ and this region. The field amplitude $|E_{\text{1D}}(z)|$ within the PhC layer can be approximated as $\sqrt{\Gamma_{\text{PhC}}/t_\text{PhC}}$, so that both integrals in Eq.~\ref{eq.10} scale proportionally to $\Delta\varepsilon_\text{PhC}\Gamma_{\text{PhC}}$. As a result, the 2D coupling coefficient consists of summation terms sharing the common factor $\Delta\varepsilon_\text{PhC}^2\Gamma_{\text{PhC}}$.

In the I-EIM method, the situation is different. There, the coefficients $\xi_{m,n}$ are independent of $z$, and the material contrast is effectively averaged over the entire multilayer stack. As a result, instead of the localized dependence $\xi_{m,n}(z\in \text{PhC})\sim\Delta\varepsilon_\text{PhC}$, one obtains a uniform coefficient $\xi_{m,n}\sim\Gamma_{\text{PhC}}\Delta\varepsilon_\text{PhC}$, since the PhC layer contributes only a fraction $\Gamma_{\text{PhC}}$ to the averaged contrast of the dielectric constant. Consequently, the integrals in Eq.~\ref{eq.9} extend over the entire PCSEL structure, yielding $\zeta^{(0,1)}_{y,m,n}\sim\Delta\varepsilon_\text{PhC}\Gamma_{\text{PhC}}$, as in the 3D-CWT case. 

However, $\kappa_{\text{2D}+}$ contains additional prefactors $\xi_{1-m,\,-n}\sim\Gamma_{\text{PhC}}\Delta\varepsilon_\text{PhC}$, so that each term in the summation for the 2D coupling coefficient scales as  $\sim\Gamma_{\text{PhC}}^2\Delta\varepsilon_\text{PhC}^2$. Thus, in contrast to 3D-CWT, an extra factor of $\Gamma_{\text{PhC}}$ appears, which must be removed by an appropriate renormalization. The same reasoning applies to $\kappa_{\text{2D}-}$, but not to $\kappa_{\text{1D}}$, whose structure is analogous to $\zeta^{(0,1)}_{y,m,n}$, but lacks an additional prefactor of the form $\xi_{m,n}$. Therefore, the issue of the redundant $\Gamma_{\text{PhC}}$ factor does not arise for the one-dimensional coupling coefficient.
\bibliography{aipsamp.bib}

\end{document}